\documentclass{article}
\usepackage{epsfig}
\usepackage{amssymb}
\newcommand{\bfr}{\begin{flushright}}
\newcommand{\efr}{\end{flushright}}
 
\begin{document}
\title{Exact solutions for gravitational collapse with a dilaton field
in arbitrary dimensions
}
\author{Takuya Maki\\
Institute of Physics, Kitasato University,\\
 Kitasato, Sagamihara-shi,
Kanagawa 228, Japan\\
and\\
Kiyoshi Shiraishi\\
Akita Junior College, Shimokitade-Sakura, Akita-shi, Akita 010, Japan
}
\date{Class. Quantum Grav. {\bf 12} (1995) pp.~159--172
}
\maketitle
\begin{abstract}
We present time-dependent analytic solutions to the Einstein equations
coupled with a dilaton (scalar) field. The background geometry for the
solutions is a product of an $N$-dimensional spherically symmetric space
and a $d$-dimensional flat space. We discuss the global properties of
the spacetime. 
\end{abstract}

\section{Introduction}
In recent years various types of inhomogeneous cosmological solutions
to the Einstein equations coupled with other fields have been studied
extensively \cite{1,2}. Such exact solutions are important for
studying the structure and properties of horizons, because knowledge of
the global structure of the spacetime is essential for such a study.

On the other hand, time dependent solutions to the Einstein
equations coupled to matter fields have been investigated as models
of gravitational collapse \cite{3}. A convenient model may involve a
scalar field as a matter field, which is coupled to Einstein gravity.
The gravitation theory including scalar fields is preferred not only
because of its simplicity, but also because it can be considered as a
reduced system of supergravity theory \cite{4} or superstring theory
\cite{5}.

Exact solutions to the coupled Einstein massless scalar field equations
have been obtained for some simple cases, including a static case
\cite{6}.  Besides a homogeneous cosmological solution, only a few
time-dependent exact solutions describing an inhomogeneous spatial
metric are known. The first example of such a solution has been given
by Roberts \cite{7}. Another type of exact solution has recently been
given by Husain, Martinez and N\'u\~nez \cite{8}.

In this paper we will obtain the time-dependent spherically symmetric
exact solution to the Einstein scalar theory in {\it arbitrary}
spacetime dimensions. This type of solution is a generalization of the
one found by Husain, Martinez and N\'u\~nez \cite{8}.%
\footnote{It is difficult to generalize Roberts' solution to the
arbitrary dimensional case.} We will consider
the metric in Kaluza-Klein pararmetrization \cite{9}, which represents
topologically a product space of a $(1+N)$-dimensional space and a
$d$-dimensional flat space. The configuration of the scalar field and
the metric is assumed to be spherical in the $N$ dimensional subspace.

As well as the massless case, in this paper we consider the case with a
potential term represented by an exponential function of the scalar
field variable. Such a potential often arises from effective field
theories of string theory or supergravity theory. In these theories the
scalar field is known as a dilaton field \cite{4,5}. The exponential
potential plays a crucial role in the exact multi-centred solution
\cite{2} as well as in exact solutions for cosmological inflation
\cite{10}.

One motivation for studying the time dependent solutions to the coupled
Einstein scalar system is that they may be regarded as an analytic
model for gravitational collapse induced by scalar fields. Recently
there has been much progress in the numerical study of
gravitational collapse. Numerical results illustrate that the
behaviour of a scalar field configuration may have a certain
self-similarity and the critical mass of the resulting black hole
may take a form governed by a certain power law with a universal
critical exponent \cite{11}. Several authors have made efforts to
understand the critical phenomena qualitatively by using the exact
solutions. Brady \cite{12} and Oshiro et al \cite{13} have discussed
the critcal exponent, based on he exact solution given by Roberts
\cite{7}, while Husain et al based their discussion on the exact
solution derived by themselves \cite{8}.

Throughout this paper we consider the action, including a dilaton
field $\phi$, of the form:
\begin{equation}
I=\int d^Dx\frac{\sqrt{-g}}{16\pi}\left[R-\frac{4}{D-2}(\nabla\phi)^2
-e^{4a\phi/(D-2)}\Lambda\right]+\mbox{boundary terms}
\label{1.1} 
\end{equation}
where $D=1+N+d$, and the Newton constant is set to unity.
Here we assume that the dilaton coupling $a$ may take an arbitrary
positive value: for effective field theories of string theory it takes
$a=1$. For $a=0$, $\Lambda$ simply becomes a cosmological constant.

In the following section we first consider the $\Lambda=0$ case, where
the dilaton field $\phi$ is a free massless field. We will treat the
case with $\Lambda\ne 0$ in section 3. In section 4 we discuss the
structure of singularities and apparent horizons in the spacetime
described by the exact solutions obtained in sections 2 and 3. In
section 5 we evaluate the mass for the self-gravitating system
described by exact solutions with the massless scalar field obtained in
section 2.

\section{Solutions for a massless scalar field}
For $\Lambda=0$ the field equations derived from the action (\ref{1.1})
take the following form:
\begin{equation}
\Box\, \phi=0
\label{2.1}
\end{equation}
\begin{equation}
R_{MN}=\frac{4}{D-2}\nabla_M\phi\nabla_N\phi\,.
\label{2.2}
\end{equation}

We wish to find the time-dependent solution to the equations (\ref{2.1})
and (\ref{2.2}) which can be interpreted as a product of an
$N$-dimensional spherically symmetric space and a $d$-dimensional flat
space.%
\footnote{For static cases several exact solutions are obtained in
\cite{14}.}
We assume that the metric should take a block diagonal form
in $d$- and
$N$-dimensional spaces. Throughout this paper we use the following
metric; 
\begin{equation}
ds^2
=\sigma^{-2/(d+1)}(r)\,[-\Delta(r)dt^2+T^2(t)d\vec{x}^2]+
\sigma^{2/(N-2)}(r)\,S^2(t)\left[\frac{dr^2}{\Delta(r)}+r^2
d\Omega^2_{N-1}\right]
\label{2.3} 
\end{equation}
where $d\vec{x}^2\equiv\sum_{i=1}^d dx_i^2$, and $d\Omega_{N-1}^2$
represents the line element of a unit ($N-1$)-sphere. Here the scale
factors $S$ and $T$ are assumed to be functions of $t$, while $\Delta$
and $\sigma$ are functions of $r$.

Now we can transform the equations (\ref{2.1}) and (\ref{2.2}) into
simultaneous differential equations concerning the unknown functions
$S$, $T$, $\Delta$, $\sigma$ and $\phi$, by taking the metric ansatz
(\ref{2.3}). Furthermore, we require the separation of equations with
the variables $t$ and $r$ to obtain analytic solutions. To this end we
will take another ansatz for the scalar field: the time derivative of
$\phi$ depends only on $t$, while the $r$-derivative of $\phi$ depends
only on $r$. We also note that the equations for the scale factors
should coincide with those known in the Kaluza-Klein cosmological
scenario \cite{4,9}.

Consequently, the reduced equations we obtain are divided into three
groups. One group includes the equations for the temporal evolution
of $S$ and $T$:
\begin{equation}
\frac{1}{S^NT^d}(S^NT^d\dot{\phi})^{\cdot}=0
\label{2.4}
\end{equation}
\begin{equation}
N\frac{\ddot{S}}{S}+d\frac{\ddot{T}}{T}=-\frac{4}{D-2}\dot{\phi}^2
\label{2.5}
\end{equation}
\begin{equation}
\frac{\ddot{S}}{S}+(N-1)\left(\frac{\dot{S}}{S}\right)^2+
d\frac{\dot{S}}{S}\frac{\dot{T}}{T}=0
\label{2.6}
\end{equation}
\begin{equation}
\frac{\ddot{T}}{T}+(d-1)\left(\frac{\dot{T}}{T}\right)^2+
N\frac{\dot{S}}{S}\frac{\dot{T}}{T}=0
\label{2.7}
\end{equation}
where a dot denotes derivation with respect to $t$.

Another group includes the equations for the spatial configuration of
$\Delta$ and $\sigma$: 
\begin{equation}
\frac{1}{r^{N-1}}(\Delta\,r^{N-1}\phi')'=0
\label{2.8}
\end{equation}
\begin{equation}
\left(\frac{1}{2}\frac{\Delta'}{\Delta}-\frac{1}{d+1}
\frac{\sigma'}{\sigma}\right)'+\left(\frac{\Delta'}{\Delta}+
\frac{N-1}{r}\right)\left(\frac{1}{2}\frac{\Delta'}{\Delta}-\frac{1}{d+1}
\frac{\sigma'}{\sigma}\right)=0
\label{2.9}
\end{equation}
\begin{eqnarray}
& &\left(\frac{1}{2}\frac{\Delta'}{\Delta}-
\frac{\sigma'}{\sigma}\right)'+\left(\frac{\Delta'}{\Delta}-
\frac{D-2}{(d+1)(N-2)}\frac{\sigma'}{\sigma}\right)\left(\frac{1}{2}\frac{\Delta'}{\Delta}-
\frac{\sigma'}{\sigma}\right)+\frac{1}{2}\frac{d}{d+1}
\frac{\Delta'}{\Delta}\frac{\sigma'}{\sigma}\nonumber \\
&
&+(N-1)\left[\left(\frac{1}{r}+\frac{1}{N-2}\frac{\sigma'}{\sigma}
\right)'+\left(\frac{1}{r}+\frac{1}{2}\frac{\Delta'}{\Delta}
\right)\left(\frac{1}{r}+\frac{1}{N-2}\frac{\sigma'}{\sigma}
\right)\right]\nonumber \\
& &=-\frac{4}{D-2}(\phi')^2
\label{2.10}
\end{eqnarray}
\begin{equation}
\left(\frac{\sigma'}{\sigma}\right)'+\left(\frac{\Delta'}{\Delta}+
\frac{N-1}{r}\right)
\frac{\sigma'}{\sigma}=0
\label{2.11}
\end{equation}
\begin{equation}
\frac{1}{r}\left[\frac{\Delta'}{\Delta}+\frac{N-2}{r}
\frac{\Delta-1}{\Delta}\right]+\frac{1}{N-2}\left[
\left(\frac{\sigma'}{\sigma}\right)'+\left(\frac{\Delta'}{\Delta}+
\frac{N-1}{r}
\right)\frac{\sigma'}{\sigma}\right]=0
\label{2.12}
\end{equation}
where the prime denotes derivation with respect to $r$.

The third group contains one equation, which involves both the time
derivative and the radial derivative of the field variables:
\begin{equation}
\frac{\dot{S}}{S}\left(\frac{N-1}{2}\frac{\Delta'}{\Delta}-
\frac{D-2}{d+1}\frac{\sigma'}{\sigma}\right)+\frac{d}{2}
\frac{\dot{T}}{T}\frac{\Delta'}{\Delta}=\frac{4}{D-2}\dot{\phi}\phi'\,.
\label{2.13}
\end{equation}

This equation (\ref{2.13}) comes from the ($rt$)-component of the
Einstein equations.

The equations (\ref{2.8}--\ref{2.12}) have a solution
\begin{equation}
\Delta(r)=1-\left(\frac{r_0}{r}\right)^{N-2}
\label{2.14}
\end{equation}
\begin{equation}
\sigma(r)=\left(\Delta(r)\right)^\gamma
\label{2.15}
\end{equation}
\begin{equation}
\phi'=\pm\frac{1}{2}(D-2)\sqrt{\frac{\gamma(1-\gamma)}{(d+1)(N-2)}}
\frac{\Delta'}{\Delta}
\label{2.16}
\end{equation}
where $r_0$ is an integration constant, and the constant $\gamma$ can
take any value in the range $0<\gamma<1$, as long as only the equations
(\ref{2.8}--\ref{2.12}) are taken into consideration.

On the other hand, the equations for the scale factors can be solved by
assuming a power law behaviour of the scale factors such as
\begin{equation}
S\propto t^\alpha\qquad \mbox{and} \qquad  T\propto t^\beta
\label{2.17}
\end{equation}
where $\alpha$ and $\beta$ are constants. Furthermore, if we take the
time derivative of the scalar field as
\begin{equation}
\dot{\phi}=\frac{\delta}{t}
\label{2.18}
\end{equation}
where $\delta$ is a constant, then the equations (\ref{2.4}--\ref{2.7})
yield the following relations among $\alpha$, $\beta$ and $\delta$:
\begin{equation}
N\alpha+d\beta=1
\label{2.19}
\end{equation}
\begin{equation}
N\alpha^2+d\beta^2+\frac{4}{D-2}\delta^2=1\,.
\label{2.20}
\end{equation}

Apparently, these relations are a generalization of the Kasner
condition in higher dimensions.

Finally, the equation (\ref{2.13}) gives a relation among $\alpha$,
$\beta$, $\gamma$ and $\delta$. We find $\delta$ can be solved in terms
of $\alpha$ and $\gamma$ and obtain
\begin{equation}
\delta=\pm\frac{1}{2}\sqrt{\frac{(d+1)(N-2)}{\gamma(1-\gamma)}}
\left(\frac{1-\alpha}{2}-\frac{D-2}{d+1}\alpha\gamma\right)  
\label{2.21}
\end{equation}
where the sign in the right-hand side matches that in equation
(\ref{2.16}). The scalar field configuration can be expressed, for
instance, as
\begin{eqnarray}
\phi&=&\phi_0\pm\frac{1}{2}(D-2)\Big[
\sqrt{\frac{\gamma(1-\gamma)}{(d+1)(N-2)}}\ln(\Delta)\nonumber \\
&
&+\sqrt{\frac{(d+1)(N-2)}{\gamma(1-\gamma)}}\left(\frac{1-\alpha}{2(D-2)}-\frac{1}{d+1}\alpha\gamma\right)
\ln\left(\frac{t}{t_0}\right)\Big]
\label{2.22}
\end{eqnarray}
where $\phi_0$ and $t_0$ are integration constants.

Since there are three equations (\ref{2.19}), (\ref{2.20}) and
(\ref{2.21}), there is only one degree of freedom in choosing the values
of the constants. For a given value for a constant, say $\gamma$
($0<\gamma<1$), the values of the other constants are given by roots of
quadratic equations in general.

Now we examine two special cases, for $d=0$ and for $N=3$.

Case 1. $d=0$ ($D=N+1$). In this case the metric does not include
the extra scale factor ($T$); therefore the values for the constants
$\alpha$, $\gamma$ and $\delta$, which appear in general cases, are
fixed as:
\begin{equation}
\alpha=\frac{1}{N}
\label{2.23}
\end{equation}
\begin{equation}
\gamma=\frac{1}{2}\pm\sqrt{\frac{N}{8(N-1)}}
\label{2.24}
\end{equation}
\begin{equation}
\delta=\pm\frac{N-1}{2\sqrt{N}}
\label{2.25}
\end{equation}
where the sign in equation (\ref{2.25}) is independent of the sign in
(\ref{2.24}).

Now the spherically symmetric metric takes the form
\begin{equation}
ds^2=-\sigma^{-2}(r)\Delta(r)dt^2+\sigma^{2/(N-2)}(r)
S^2(t)\left[\frac{dr^2}{\Delta(r)}+r^2d\Omega^2_{N-1}\right]
\label{2.26}
\end{equation}
where
\begin{equation}
\Delta(r)=1-\left(\frac{r_0}{r}\right)^{N-2}
\label{2.27}
\end{equation}
\begin{equation}
\sigma(r)=(\Delta(r))^\gamma\qquad\mbox{with}\qquad\gamma=\frac{1}{2}
\pm\sqrt{\frac{N}{8(N-1)}}
\label{2.28}
\end{equation}
\begin{equation}
S\propto t^{1/N}\,.
\label{2.29}
\end{equation}

The scalar field is then
\begin{equation}
\phi=\phi_0\pm\frac{1}{2}\left[\sqrt{\frac{1}{8}(N-1)}\ln(\Delta)
+\mbox{sign}\left(\frac{1}{2}-\gamma\right)\frac{N-1}{\sqrt{N}}
\ln\left(\frac{t}{t_0}\right)\right]\,.
\label{2.30}
\end{equation}

Case 2. $N=3$ ($D=d+4$). In this case the value for the constants
$\alpha$, $\beta$, $\gamma$ and $\delta$ are determined by solving the
equation for $\alpha$ and $\gamma$
\begin{equation}
3\alpha^2+\frac{(1-3\alpha)^2}{d}+\frac{1}{\gamma(1-\gamma)}
\frac{d+2}{d+1}\left(\frac{d+1}{d+2}
\frac{1-\alpha}{2}-\alpha\gamma\right)^2=1
\label{2.31}
\end{equation}
and the other equations yield the values for $\beta$ and $\delta$. The
contours for possible values for $\alpha$, and $\gamma$ are plotted in
figure 1, where the contours for $d=1$ and for $d=\infty$ are shown. As
seen from figure 1, we can find a set of parameters leaving one degree
of freedom for $d\ne 0$, in general.

\begin{figure}[ht]
\begin{center}
\includegraphics[width=6cm]{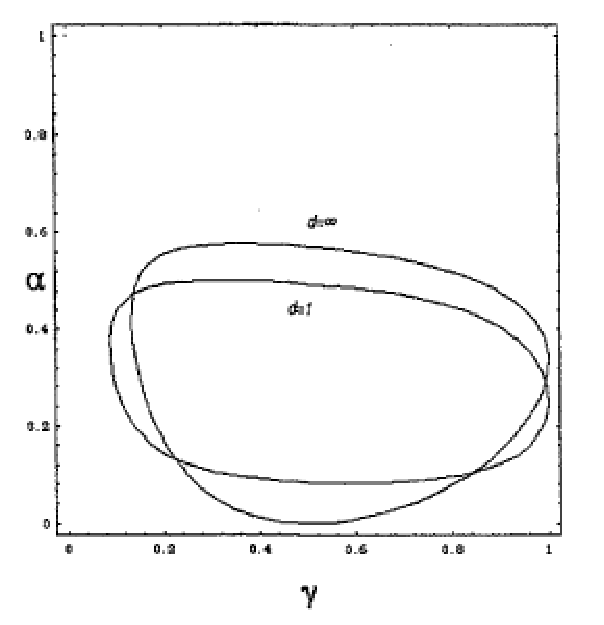}
\caption{The contours for possible values for $\alpha$ and $\gamma$ in
the exact solutions obtained for a massless scalar field and
$N=3$ (in section 2); the contours are for $d=1$ and for $d=\infty$, as
indicated.}
\label{f1}\end{center}
\end{figure}

In the following section we consider the exact solution to the
system governed by the action (\ref{1.1}) with non-zero $\Lambda$.

\section{Solutions for a scalar field with an exponential potential}
Now we turn to the $\Lambda\ne 0$ case. The field equations then take
die following form:
\begin{equation}
\Box\phi=\frac{1}{2}a\Lambda e^{4a\phi/(D-2)}
\label{3.1}
\end{equation}
\begin{equation}
R_{MN}=\frac{4}{D-2}\nabla_M\phi\nabla_N\phi+\frac{\Lambda}{D-2}
e^{4a\phi/(D-2)}g_{MN}\,.
\label{3.2}
\end{equation}

We use the same ans\"atze for the metric and the dilaton scalar
field as in the previous section. To obtain analytic solutions we assume
that the equations (\ref{2.8}--\ref{2.13}) are unchanged even if there
is a potential term. This assumption with the homogeneous cosmological
solution \cite{10}. 

We find that the assumption requires two constraints
on the field variables as follows:
\begin{equation}
T=S\qquad\mbox{and}\qquad\left(\frac{\Delta}{\sigma^{2/(d+1)}}
e^{4a\phi/(D-2)}\right)'=0\,.  
\label{3.3}
\end{equation}

The differential equations including the scale factor, which
correspond to (\ref{2.4}--\ref{2.7}), are reduced to:
\begin{equation}
\ddot{\phi}+(D-1)\frac{\dot{S}}{S}\dot{\phi}=-\frac{1}{2}a\Lambda
\left(\frac{\Delta}{\sigma^{2/(d+1)}} e^{4a\phi/(D-2)}\right)
\label{3.4}
\end{equation}
\begin{equation}
(D-1)\frac{\ddot{S}}{S}=-\frac{4}{D-2}\dot{\phi}^2+\frac{\Lambda}{D-2}
\left(\frac{\Delta}{\sigma^{2/(d+1)}} e^{4a\phi/(D-2)}\right)
\label{3.5}
\end{equation}
\begin{equation}
\frac{\ddot{S}}{S}+(D-2)\left(\frac{\dot{S}}{S}\right)^2
=\frac{\Lambda}{D-2}
\left(\frac{\Delta}{\sigma^{2/(d+1)}} e^{4a\phi/(D-2)}\right)\,.
\label{3.6}
\end{equation}
The equation (\ref{2.14}) becomes
\begin{equation}
(D-2)\frac{\dot{S}}{S}\left(\frac{1}{2}\frac{\Delta'}{\Delta}-
\frac{1}{d+1}\frac{\sigma'}{\sigma}\right)=\frac{4}{D-2}\dot{\phi}\phi'
\label{3.7}
\end{equation}

The differential equations (\ref{2.8}--\ref{2.13}) remain vaild in this
case. Thus they again call for the functions of $r$ as
\begin{equation}
\Delta(r)=1-\left(\frac{r_0}{r}\right)^{N-2}
\label{3.8}
\end{equation}
\begin{equation}
\sigma(r)=(\Delta(r))^\gamma
\label{3.9}
\end{equation}
\begin{equation}
\phi'=\pm\frac{1}{2}\sqrt{\frac{\gamma(1-\gamma)}{(d+1)(N-2)}}
\frac{\Delta'}{\Delta}
\label{3.10}
\end{equation}
where $\gamma$ is left undetemined in this step.

As long as the metric (\ref{2.3}) is used, the solutions for $S$ are
treated separately for $a\ne 0$ and for $a=0$.

For $a\ne 0$ we have the following solutions to (\ref{3.3}--\ref{3.6})
\begin{equation}
S\propto t^{1/a^2}
\label{3.11}
\end{equation}
\begin{equation}
\dot{\phi}=-\frac{D-2}{2a}\frac{1}{t}
\label{3.12}
\end{equation}
with
\begin{equation}
\frac{\Delta}{\sigma^{2/(d+1)}}e^{4a\phi/(D-2)}=
\frac{(D-2)(D-1-a^2)}{\Lambda a^4}\frac{1}{t^2}
\label{3.13}
\end{equation}

Note that the sign of the time derivative of $\phi$ is given
definitely in this case ($\Lambda\ne 0$).

Substituting (\ref{3.8}--\ref{3.12}) into (\ref{3.7}), we find that
$\gamma$ is a root of the following quadratic equation
\begin{equation}
\frac{a^2(d+1)\gamma(1-\gamma)}{N-2}=\left(\frac{1}{2}(d+1)-\gamma
\right)^2
\label{3.14}
\end{equation}
and the scalar field is expressed as
\begin{equation}
\phi=\phi_0-\frac{D-2}{4a}\left(1-\frac{2\gamma}{d+1}\right)\ln(\Delta)
-\frac{D-2}{2a}\ln\left(\frac{t}{t_0}\right)\qquad(a\ne 0)
\label{3.15}
\end{equation}
where $\phi_0$ and $t_0$ are constants that are mutually related through
the relation (\ref{3.13}). The equation (\ref{3.14}) has real solutions
if and only if $a^2\ge (N-2)(d-1)$. This constrains the possible range
of $a^2$ if $d\ge 2$.

Turning to the case with $a=0$, we have the solutions to
(\ref{3.3}--\ref{3.6}):
\begin{equation}
S\propto e^{Ht}
\label{3.16}
\end{equation}
with
\begin{equation}
H^2=\frac{\Lambda}{(D-1)(D-2)}
\label{3.17}
\end{equation}
and
\begin{equation}
\dot{\phi}=0\,.
\label{3.18}
\end{equation}

In addition it is required that
\begin{equation}
\gamma=\frac{1}{2}(d+1)
\label{3.19}
\end{equation}
which comes from (\ref{3.3}). For $a=0$ we obtain the exact solution
only for $d=0$ and for $d=1$.

\begin{itemize}
\item $d =0$. Then $\gamma=\frac{1}{2}$. The scalar field is expressed
as
\begin{equation}
\phi=\phi_0\pm\frac{D-2}{4\sqrt{N-2}}\ln(\Delta)
\label{3.20}
\end{equation}
where $\phi_0$ is a constant.

The metric then takes the form:
\begin{equation}
ds^2=-dt^2+\Delta^{1/(N-2)}(r)
S^2(t)\left[\frac{dr^2}{\Delta(r)}+r^2d\Omega_{N-1}^2\right]\,.
\label{3.21}
\end{equation}

\item $d=1$. Then $\gamma=1$. The scalar field is constant everywhere,
\begin{equation}
\phi=\phi_0\,.
\label{3.22}
\end{equation}

The metric then looks like
\begin{equation}
ds^2=-dt^2+S^2(t)\left\{\frac{dx^2}{\Delta(r)}+\Delta^{2/(N-2)}(r)
\left[\frac{dr^2}{\Delta(r)}+r^2d\Omega_{N-1}^2\right]\right\}\,.
\label{3.23}
\end{equation}
\end{itemize}

The global structure of the spacetime expressed by the exact solutions
will be examined in the subsequent section.

\section{The global structure of the spacetime}
We examine the global property of the spacetime described by the
solutions obtained in sections 2 and 3. Here we take the
$d$-dimensional space as an extra space, and so we study the
($N+1$)-dimensional spacetime as a physical spacetime. Therefore we
will concentrate attention on the time and radial part of the metric.
Furthermore, it is useful to transform the time coordinate $t$ into
$\eta$, which satisfies
\begin{equation}
d\eta^2=\frac{dt^2}{S^2(t)}\,.
\label{4.1}
\end{equation}

Then the ($\eta, r$) part of the metric is expressed as
\begin{equation}
ds^2=S^2(\eta)\left[-\sigma^{-2/(d+1)}(r)\Delta(r)d\eta^2+
\frac{\sigma^{2/(N-2)}(r)}{\Delta(r)}dr^2\right]\,.
\label{4.2}
\end{equation}

Here we must recall
\begin{equation}
\Delta(r)=1-\left(\frac{r_0}{r}\right)^{N-2}\qquad\mbox{and}\qquad
\sigma(r)=(\Delta(r))^\gamma
\label{4.3}
\end{equation}
where $r_0$ is an arbitrary constant.

Using this `conformally invariant time coordinate' $\eta$, the scale
factor $S$ in the case with $\Lambda=0$, treated in section 2, can be
written as
\begin{equation}
S(\eta)=(A\eta+B)^{\alpha/(1-\alpha)}
\label{4.4}
\end{equation}
where $A$ and $B$ is arbitrary constants. We remember that there is a
rehtion between $\alpha$ ard $\gamma$, which can be obtained from
(\ref{2.19}--\ref{2.21}),
\begin{equation}
N\alpha^2+\frac{(1-N\alpha)^2}{d}+\frac{(d+1)(N-2)}{(D-2)
\gamma(1-\gamma)}\left(\frac{1}{2}(1-\alpha)-
\frac{D-2}{d+1}\alpha\gamma\right)^2=1\,.
\label{4.5}
\end{equation}

For the $\Lambda\ne 0$ treated in section 3, the relations among the
several constants take different forms. In the coordinate system that
includes the conformally invariant time, the expressions for the
solutions take dissimilar forms for $a\ne 1$ and for $a=1$.

For $a\ne 1$, the solutions can be written as
\begin{equation}
S(\eta)=(A\eta+B)^{-1/(1-a^2)}
\label{4.6}
\end{equation}
\begin{equation}
\phi=\phi_0-\frac{D-2}{4a}\left(1-\frac{2\gamma}{d+1}\right)\ln(\Delta)
+\frac{(D-2)a}{2(1-a^2)}\ln(A\eta+B)
\label{4.7}
\end{equation}
with
\begin{equation}
\frac{a^2(d+1)\gamma(1-\gamma)}{N-2}=\left(\frac{1}{2}(d+1)-
\gamma\right)^2\qquad(0<\gamma<1)
\label{4.8}
\end{equation}
\begin{equation}
e^{4a\phi_0/(D-2)} =\frac{(D-2)(D-1-a^2)}{\Lambda(1-a^2)^2} A^2
\label{4.9}
\end{equation}
where $A$, $B$ and $\phi_0$ are constants. The value for $a^2$ must be
greater than $(N-2)(d-1)$, for the equation (\ref{4.8}) has real roots.

For $a=1$ we find the form:
\begin{equation}
S(\eta)=e^{h(\eta-\eta_0)}
\label{4.10}
\end{equation}
\begin{equation}
\phi=\phi_0-\frac{1}{4}(D-2)\left(1-\frac{2\gamma}{d+1}\right)\ln(\Delta)
+\frac{1}{2}(D-2)h(\eta-\eta_0)
\label{4.11}
\end{equation}
with
\begin{equation}
\frac{(d+1)\gamma(1-\gamma)}{N-2}=\left(\frac{1}{2}(d+1)-\gamma
\right)^2\qquad(0<\gamma<1)
\label{4.12}
\end{equation}
\begin{equation}
e^{4\phi_0/(D-2)}=\frac{(D-2)^2}{\Lambda}h^2
\label{4.13}
\end{equation}
where $h$, $\eta_0$ and $\phi_0$ are constants. Incidentally, the
equation (\ref{4.12}) has real roots only for $d=0,1$ or $d=2$ and
$N=3$.

The structure of the spacetime singularity involved in this exact
solution is as follows: the timelike singulality is located at $r=r_0$,
regardless of the presence of $\Lambda$ and of the value for $a$.%
\footnote{This is guaranteed by the fact that $[(D-2)/((N-2)(d+
1))]\gamma<1$.}
However, the property of another `cosmological' singularity depends on
the value for $\Lambda$ and $a$.

Let us consider the contracting universe, i.e. the case with the scale
factor decreasing with time. This corresponds to choosing $A<0$ in
(\ref{4.4}) and $A>0$ in (\ref{4.6}), and $h<0$ in (\ref{4.10}). We
further assume that the values of constants $B$ and $\eta_0$ are zero.

For $\Lambda=0$ the spacelike singularity is located at $\eta=0$, which
is just the `big crunch' singularity. The range of the cosmological
time is $-\infty<\eta<0$. The spacelike singularity lies at $\eta=0$.
Inward going radial light rays reach either the timelike or the
spacelike singularity, while any outward going radial light rays reach
the null singularity.

On the other hand, for $\Lambda\ne 0$, the global structure of the
singularities depends on the value of $a$ (see figure 2).

\begin{figure}[ht]
\begin{center}
\includegraphics[width=6cm]{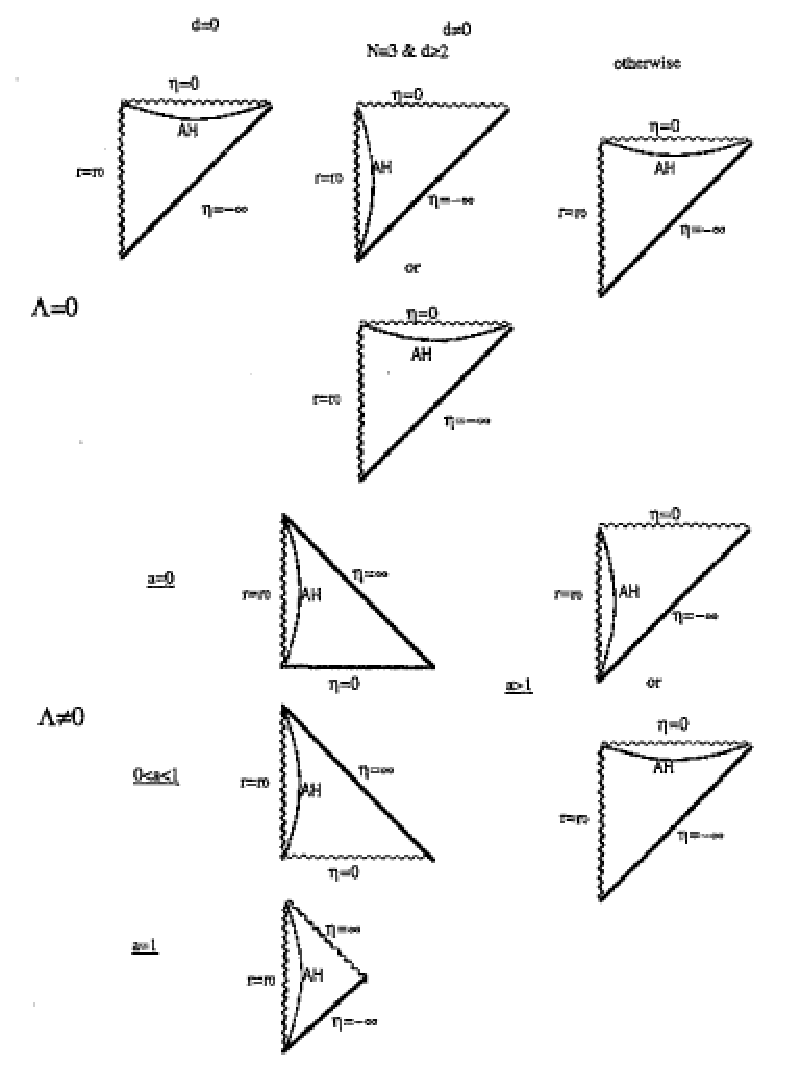}
\caption{The schematic view of the global property of the
spacetime described by the exact solutions obtained in sections 2 and
3.}
\label{f2}\end{center}
\end{figure}

For $a>1$ the global structure of the singularities is the same as
the case of $\Lambda= 0$. The spacelike singularity lies at $\eta=0$
(the range of $\eta$ is $-\infty<\eta<0$).

For $a=1$ there is a null singularity at $\eta=\infty$ (the range of
$\eta$ is $-\infty<\eta<\infty$) \cite{15}. Inward going light rays hit
the timelike singularity at $r=r_0$, while any outward going light
rays reach the null singlarity.

For $0<a<1$ the cosmological singularity is still null. But the
range of the cosmic time is $0<\eta<\infty$.

For $a=0$ there is no other singularity besides the timelike one at
$r=r_0$, because the spacetime is asymptotically de Sitter space in
this case.

Next we examine apparent horizons in these spacetimes. It is
significant to study the property of apparent horizons, especially
when the metric is not static. The hiding of singularities by apparent
horizons may imply the formation of black holes in many cases, for
example.

An apparent horizon is lhe surface defined by
\begin{equation}
g^{MN}R,_MR,_N=0
\label{4.14}
\end{equation}
where
\begin{equation}
R(\eta, r)\equiv S(\eta) r \sigma^{1/(N-2)}(r)\,.
\label{4.15}
\end{equation}

In our case the equation (\ref{4.14}) leads to
\begin{equation}
\frac{1}{S}\frac{dS}{d\eta}=\pm\frac{1}{r}\left[1-(1-\gamma)
\left(\frac{r_0}{r}\right)^{N-2}\right]
\left[1-\left(\frac{r_0}{r}\right)^{N-2}\right]^{-\frac{D-2}{(N-2)(d+1)}\gamma} 
\label{4.16}
\end{equation}

This defines the apparent horizon in the spacetime.

The apparent horizon may be spacelike, null, or timelike in
different regions in general cases. The property can be indicated by
the ratio of the slope of the apparent borizon and the null ray
\cite{8}. The absolute value of the ratio is given in our case as
\begin{equation}
\left|\frac{d\eta_{AH}}{d\eta_{Null}}\right|=
\left|\frac{\alpha}{1-\alpha}\right|\left\{\frac{D-2}{d+1}-
\frac{N-2}{d+1}\frac{[1-(r_0/r)^{N-2}][1+d(1-\gamma)
(r_0/r)^{N-2}]}{[1-(1-\gamma)(r_0/r)^{N-2}]^2}
\right\} 
\label{4.17}
\end{equation}
for
the case with massless scalar. For the case with the exponential scalar
potential ($a\ne 1$) equation (\ref{4.17}) is still valid if $\alpha$
in the equation be replaced by $1/a^2$. For $a=1$ equation (\ref{4.16})
merely determines the location of the timelike apparent horizon. Now,
let us examine the value of (\ref{4.17}) for each exact solution
obtained in sections 2 and 3.

First we examine the massless scalar case, treated in section 2. For
$d=0$, i.e. thee is no `extra' dimension other than the
spherically symmetric space, the equation (\ref{4.17}) then becomes
\begin{equation}
\left|\frac{d\eta_{AH}}{d\eta_{Null}}\right|=1-\frac{N-2}{N-1}
\frac{1-(r_0/r)^{N-2}}{1-(1-\gamma)(r_0/r)^{N-2}}<1
\qquad(\mbox{for~} d= 0, \mbox{massless}) 
\label{4.18}
\end{equation}
since the value of $\alpha$, has been uniquely solved as $\alpha=1/N$
(\ref{2.23}). In this case die apparent horizon is spacelike for
$r>r_0$. Husain et al found this feature for the $N=3$ case \cite{8},
and we find there that the same characteristic of the apparent horizon
holds for arbitrary dimensions of spherically symmetric space. For $d\ge
1$ the apparent horizon may be spacelike, null, or timelike in different
regions, in general. We can, however, and a set or parameters ($\alpha$
and
$\gamma$) which allows a spacelike apparent horizon in all regions. One
cannot choose a set of the pammeters that admits a timelike apparent
hoizon in all of the spacetime region, except for $N=3$ and $d\ge 2$.

Next we turn to the case with $\Lambda\ne 0$. The property of the
apparent horizon depends on $a$ as well as $N$ and $d$. It is classified
into four cases:

\begin{itemize}
\item $a=0$. To require an exact solution of the type we have
considered in this paper, the allowed values for $d$ are $d=0$ and
$d=1$. For each case the apparent horizon is timelike, except at
$\eta=0$, $r=r_0$.

\item $0<a<1$. In this case $d=0$ and $d=1$ are permitted as well. The
apparent horizon may be timelike, null, or spacelike in a different
region. But in the viciniy of $r=r_0$ and at spatial infinity the
apparent horizon is always timelike.

\item $a=1$. The allowed values for $d$ are $d=0$, $d=1$, and $d=2$.
Further, for
$d=2$ only $N=3$ is pemitted. For each case the apparent horizon is
timelike everywhere in the spacetime.

\item $a>1$. In this case any value for $d$ is permitted. but the value
for $N$ is restricted by $a$ and $d$ for $d\ge 2$ through the inequality
$a^2\ge (N-2)(d-1)$. The apparent horizon may be timelike, null, or
spacelike in a different region. For a sufficiently large $a$ the
apparent horizon becomes spacelike everywhere.
\end{itemize}

The schematic view of the global structue of the spacetime
represented by our exact solutions is exhibited in figure 2. The cases
in which the spacelike apparent horizon covers the future singularity
can be regarded as models of a gravitational collapse, provided that
the timelike singularity can be ignored.

\section{Masses for the spherical system}
In this section we evaluate the mass of the self-gravitating system of
a scalar field described by the exact solutions obtained in section 2.

In the present analysis we adopt the exact solutions with $d=0$,
which describes spherically symmetric spacetimep in this case the exact
solution can be regarded as a model of gravitational collapse, since
the future spacelike singularity is covered by the apparent horizon. The
mass can be defined by local variables in the spherically symmetric
system \cite{16}:
\begin{equation}
m(\eta, r)=\frac{(N-1)A_{N-1}}{16\pi}R^{N-2}(1-g^{MN}R,_M R,_N)
\label{5.1}
\end{equation}
where $A_{N-1}=2\pi^{N/2}/\Gamma(N/2)$ and $R$ is defined by
(\ref{4.15}).

On the apparent horizon, the mass can be written in the following
form: 
\begin{equation}
M_{AH}=\frac{(N-1)A_{N-1}}{16\pi}\left(\frac{|A|}{N-1}
\right)^{\frac{N-2}{N-1}}
\frac{r^{N(N-2)/(N-1)}[1-(r_0/r)^{N-2}]^{2\gamma}
}{[1-(1-\gamma)(r_0/r)^{N-2}]^{(N-2)/(N-1)}}\,.
\label{5.2} 
\end{equation}
This expression
exhibits a complicated form, which depends non-trivially on the
spacetime dimension.

 The exact solution obtained in section 2 may not be appropriate for a
model of gravitational collapse, because of the singular behaviour of
the scalar field configuration as an initial condition. Our result,
however, suggests that some physical quantities may depend on the
space dimension. Thus we also feet interest in numerical study of the
higher-dimensional gravitating system.%
\footnote{The authors of \cite{8,12,13,17,18} are also interested in
the analytical interpretation of the numerical results.}

\section{Conclusion}
In this paper we have given exact solutions for a spherical collapse
driven by a scalar field with an exponential potential in
$D$-dimensional spacetime. The solutions describe evolution of a scalar
field configuration in the background metric of a product of a
spherically symmetric space and an internal space. This solution
could be generalized for various supergravity models, which include
dilaton fields.

\section*{Acknowledgments}
The authors thank the referees for critical comments.


\end{document}